\documentclass[a4paper]{article}

\usepackage{INTERSPEECH2021}
\usepackage{multirow,diagbox}
\usepackage{booktabs}
\usepackage{threeparttable}
\usepackage{algorithm}
\usepackage{algorithmic}
\usepackage{amsmath}
\usepackage[colorlinks,linkcolor=blue]{hyperref}

\title{Multi-Speaker ASR Combining Non-Autoregressive Conformer CTC and Conditional Speaker Chain}
\name{Pengcheng Guo$^1$, Xuankai Chang$^2$, Shinji Watanabe$^2$, Lei Xie$^{1*}$\thanks{*Lei Xie is the corresponding author.}}
\address{
  $^1$ASLP@NPU, School of Computer Science, Northwestern Polytechnical University, Xi’an, China\\
  $^2$Carnegie Mellon University, USA}
\email{pcguo@nwpu-aslp.org, xuankaic@andrew.cmu.edu, shinjiw@ieee.org, lxie@nwpu.edu.cn}

\begin{document}

\maketitle
\begin{abstract}
Non-autoregressive (NAR) models have achieved a large inference computation reduction and comparable results with autoregressive (AR) models on various sequence to sequence tasks. However, there has been limited research aiming to explore the NAR approaches on sequence to multi-sequence problems, like multi-speaker automatic speech recognition (ASR). In this study, we extend our proposed conditional chain model to NAR multi-speaker ASR. Specifically, the output of each speaker is inferred one-by-one using both the input mixture speech and previously-estimated conditional speaker features. In each step, a NAR connectionist temporal classification (CTC) encoder is used to perform parallel computation. With this design, the total inference steps will be restricted to the number of mixed speakers. Besides, we also adopt the Conformer and incorporate an intermediate CTC loss to improve the performance. Experiments on WSJ0-Mix and LibriMix corpora show that our model outperforms other NAR models with only a slight increase of latency, achieving WERs of 22.3\% and 24.9\%, respectively. Moreover, by including the data of variable numbers of speakers, our model can even better than the PIT-Conformer AR model with only 1/7 latency, obtaining WERs of 19.9\% and 34.3\% on WSJ0-2mix and WSJ0-3mix sets. All of our codes are publicly available at \href{https://github.com/pengchengguo/espnet/tree/conditional-multispk}{https://github.com/pengchengguo/espnet/tree/conditional-multispk}.
\end{abstract}

\section{Introduction}\label{sec:intro}
End-to-end architectures have demonstrated their effectiveness and became the dominant models across various sequence to sequence tasks, like neural machine translation (NMT)~\cite{bahdanau2014neural, vaswani2017attention} and automatic speech recognition (ASR)~\cite{chan2015listen, dong2018speech, karita2019comparative, gulati2020conformer, guo2021recent}. However, most of these models follow an autoregressive (AR) strategy, which predicts a target token conditioned on both previously generated tokens and the source input sequence. The incremental process makes it hard to compute parallel and results in a large latency during the inference. In contrary to AR models, non-autoregressive (NAR) models have drawn immense interest recently, aiming to get rid of the temporal dependency and perform parallel inference.

NAR models were first proposed in NMT and have achieved competitive performance with conventional AR models~\cite{gu2017non, libovicky2018end, lee2018deterministic, stern2019insertion, gu2019levenshtein, ghazvininejad2019mask, ghazvininejad2020semi, saharia2020non}. The idea of NAR models is to predict the whole target sequence within a \textit{constant} number of iterations which is not strict with the sequence length. In~\cite{gu2017non}, Gu \textit{et al.} introduced a fertility module to predict the number of times each encoder output should be repeated and regraded the repeated encoder outputs as decoder input to perform parallel inference. In~\cite{lee2018deterministic}, Lee \textit{et al.} proposed a deterministic NAR model by iteratively refine the outputs from corrupted predictions. In addition, there were lots of studies based on the insert or edit sequence generation~\cite{stern2019insertion, gu2019levenshtein}, connectionist temporal classification (CTC)~\cite{libovicky2018end}, and masked language model objective~\cite{ghazvininejad2019mask, ghazvininejad2020semi, saharia2020non}. 

Inspired by the success of NAR models in NMT, several NAR methods were also proposed to reach the performance of AR models on ASR~\cite{chen2019listen, higuchi2020mask,chan2020imputer,tian2020spike,higuchi2020improved,chi2020align,fan2020cass}. Since CTC learns a frame-wise latent alignment between the input speech and output tokens and predicts the target sequence based on a strong conditional independence assumption~\cite{graves2006connectionist}, it can be viewed as an early-stage realization of NAR ASR models. In ~\cite{chan2020imputer}, Imputer was proposed to iteratively generate a new CTC alignment based on mask prediction. Besides, Mask-CTC~\cite{higuchi2020mask, higuchi2020improved} and Align-Refine~\cite{chi2020align} aimed to refine a token-level CTC output or latent alignments with the mask prediction. In~\cite{tian2020spike}, Tian \textit{et al.} proposed to use the estimated CTC spikes to predict the length of target sequence and adopt the encoder states as the input of decoder. However, most of aforementioned methods mainly focus on sequence to sequence tasks, like NMT and single-speaker ASR, and it is hard to directly extended to sequence to multi-sequence tasks, like multi-speaker ASR.
 
Multi-speaker ASR aims to predict the corresponding transcription for each speaker from multiple speakers overlapping speech. Although lots of AR models were explored for multi-speaker ASR, such as permutation invariant training (PIT)~\cite{qian2018single} or deep clustering (DPCL)~\cite{menne2019analysis} based hybrid system and recurrent neural network (RNN) or Transformer based end-to-end model models~\cite{seki2018purely, chang2019end, chang2020end, kanda2020joint, von2020end}, few attempts have been made to realize NAR training. 
In this study, we revisit the proposed conditional chain based methods~\cite{von2019all, shi2020speaker, shi2020sequence, fujita2020neural} and extend it to NAR multi-speaker ASR.
By doing this, the output of each speaker is predicted one-by-one by making use of both the mixed input as well as previously-estimated conditional speaker features. In each prediction step, a CTC-based NAR encoder network is used to perform parallel computation. Since the performance of CTC may suffer a severe degradation due to the conditional independence assumption, we also explore adopting an advanced Conformer encoder~\cite{gulati2020conformer} architecture to capture both local and global acoustic dependencies and an additional intermediate loss~\cite{lee2021intermediate} as a regularization function. Finally, while the original conditional chain model takes the token-level CTC alignments as the ``hard" conditional speaker features, we propose to use ``soft" conditions which are latent feature representations extracted after the last encoder layer. We evaluate the effectiveness of our model on two multi-speaker ASR benchmarks, WSJ0-Mix and LibriMix. Both results outperform other NAR models with a minor increment of latency and even achieve comparable results with the AR models.


\section{End-to-end Multi-speaker ASR}\label{sec:multispkr-asr}
We briefly introduce the end-to-end multi-speaker ASR in this section. Given the input features of the mixture speech $\mathbf{X} = \{ \mathbf{x}_1, \dots, \mathbf{x}_T \}$, where $T$ means the number of frames, the target is to directly predict the transcriptions $\mathbf{Y} = \{ \mathbf{Y}^1, \dots, \mathbf{Y}^J$ for different speakers, where $J$ refers to the number of mixed speakers. 
In~\cite{seki2018purely, chang2020end}, an end-to-end multi-speaker ASR model was proposed, which is based on the joint CTC/attention-based encoder-decoder framework~\cite{kim2017joint}. The encoder of the model consists of a mixture encoder, speaker-dependent (SD) encoders, and a recognition encoder. For a mixture speech $X$, the encoder output can be formulated as:
\begin{align}
    \mathbf{H} &= \text{Encoder}_{\text{Mix}}(\mathbf{X}) \label{eq:enc_mix}, \\
    \mathbf{H}^{j} &= \text{Encoder}_{\text{SD}}^{j}(\mathbf{H}),\, j=1,\dots,J, \label{eq:enc_sd}\\
    \mathbf{G}^{j} &= \text{Encoder}_{\text{Rec}}(\mathbf{H}^{j}),\, j=1,\dots,J. \label{eq:enc_rec}
\end{align}
Firstly, the mixture encoder in Eq.~(\ref{eq:enc_mix}) encodes the input features as hidden representations $\mathbf{H}$ of the mixture speech. Then, $J$ speaker-different encoders extract each speaker's speech representation $\mathbf{H}^j$ from the mixture representation. Next, the recognition encoder maps speech representations of each speaker into the high-level embeddings $\mathbf{G}^j$. Following the encoder, a CTC objective function is used to train the encoder and determine the best permutation $\hat{\pi}$ of the embeddings by permutation invariant training (PIT)~\cite{yu2017permutation}:
\begin{align}
    \hat{\pi} = \arg\min_{\pi \in \mathcal P} \sum_{j} \mathrm{Loss}_{\text{CTC}} (\mathbf{G}^j, \mathbf{Y}^{\pi(j)}),\; j=1,\dots,J, \label{eq:pit-ctc}
\end{align}
where $\mathbf{Y}^j$ is the $j$-th reference transcription and $\mathcal P$ means the set of all permutations on $\{ 1, \dots, J\}$.

After encoder, an attention-based decoder takes each high-level embedding $\mathbf{G}^j$ and generates the hypothesis $\widetilde{\mathbf{Y}}^j$. The computation of the decoder is:
\begin{align}
    \mathbf{c}^{j}_n &= \text{Attention} (\mathbf{e}^{j}_{n-1}, \mathbf{G}^{j}) ,\\
    \mathbf{e}^{j}_{n} &= \text{Update}(\mathbf{e}^{j}_{n-1}, \mathbf{c}^{j}_{n-1}, \widetilde{y}^{j}_{n-1}), \\
    \widetilde{y}^{j}_n &\sim \text{Decoder} (\mathbf{e}^{j}_n, \widetilde{y}^{j}_{n-1}),
\end{align}
in which $\mathbf{c}^{j}_n$ denotes the context vector and $\mathbf{e}^{j}_{n}$ is the hidden state of the decoder at step $n$. The permutation computed in the CTC step (as in Eq.~(\ref{eq:pit-ctc})) also plays an important role in the decoder, determining the order of the reference sequences for the cross-entropy (CE) loss function and the input history of teacher-forcing training. The final loss function is a combination of the CTC loss and the decoder CE loss:
\begin{equation}
    \begin{split}
    \mathcal{L} = \sum_j &\left( \lambda \mathrm{Loss}_{\text{CTC}} (\mathbf{G}^j, \mathbf{Y}^{\hat{\pi}(j)}) + \right. \\
    & \left. (1-\lambda) \mathrm{Loss}_{\text{Attn}} (\widetilde{\mathbf{Y}}^j, \mathbf{Y}^{\hat{\pi}(j)}) \right), \label{eq:tot_loss}     
    \end{split}
\end{equation}
where $\lambda$ is an interpolation factor to scale different losses.


\section{Non-autoregressive Multi-speaker ASR}\label{sec:nar-multispkr-asr}
End-to-end models proposed in Section~\ref{sec:multispkr-asr} mainly focus on an AR strategy, which will be cumbered with a complex computation and large latency problems. Although an encoder-only CTC framework can be regarded as a NAR model, the system may be susceptible to performance degradation due to the conditional independence assumption. In this study, we revisit our proposed conditional speaker chain based method for NAR multi-speaker ASR. The improved model consists of a conditional speaker chain module and Conformer CTC encoders.
While the conditional speaker chain explicitly models the relevance between outputs of different iterations, the Conformer CTC aims to conduct NAR computation in each single step. The total inference steps are restricted to the number of mixed speakers. In addition, we also explore incorporating an intermediate CTC loss as a regularization function to further improve the system performance. 

\subsection{Conformer Encoder}\label{subsec:conformer}
Conformer encoder~\cite{gulati2020conformer} is a stacked multi-block architecture, which includes a multihead self-attention (MHSA) module, a convolution (CONV) module, and a pair of positionwise feed-forward (FNN) module in the Macaron-Net style. While the MHSA learns the global context, the CONV module efficiently captures the local correlations synchronously. Since the Conformer encoder has shown consistent improvement over a wide range of end-to-end speech processing applications~\cite{guo2021recent}, we expect it to compensate for the modeling capacity of CTC and improve the system performance.

\subsection{Intermediate CTC Loss}\label{subsec:interctc}
In~\cite{lee2021intermediate}, Lee \textit{et al.} proposed a simple but efficient auxiliary loss function for CTC based ASR models, named intermediate CTC loss. The main idea of intermediate CTC loss is to choose an intermediate layer within the encoder network and induce a sub-network by skipping all higher layers after the selected layer. By computing the additional CTC loss w.r.t the output of intermediate layer, the sub-network relies more on the lower layers instead of the higher layers, which can regularize the model training. Choosing the $m$-th layer from a $L$-layer encoder network, its output can be defined as $\mathbf{H}_{m}^j$. Thus, the final loss of our model becomes:
\begin{equation}
\begin{split}
    \mathcal{L} = \sum_j & \left( (1 - \lambda) \mathrm{Loss}_{\text{CTC}} (\mathbf{G}^j, \mathbf{Y}^{\pi(j)}) + \right. \\ 
    & \left. \lambda \text{Loss}_{\text{InterCTC}} (\mathbf{H}_{m}^j, \mathbf{Y}^{\pi(j)}) \right), \label{eq:our_tot_loss}
\end{split}
\end{equation}
where $\lambda$ refers to the weight of intermediate loss. In this work, we set $\lambda$ equals to 0.1 and choose a middle layer of the $\text{Encoder}_{\text{Rec}}$ as the intermediate layer ($m = L/2$).

\begin{figure}[tbp]
    \centering
    \includegraphics[width=1.0\linewidth]{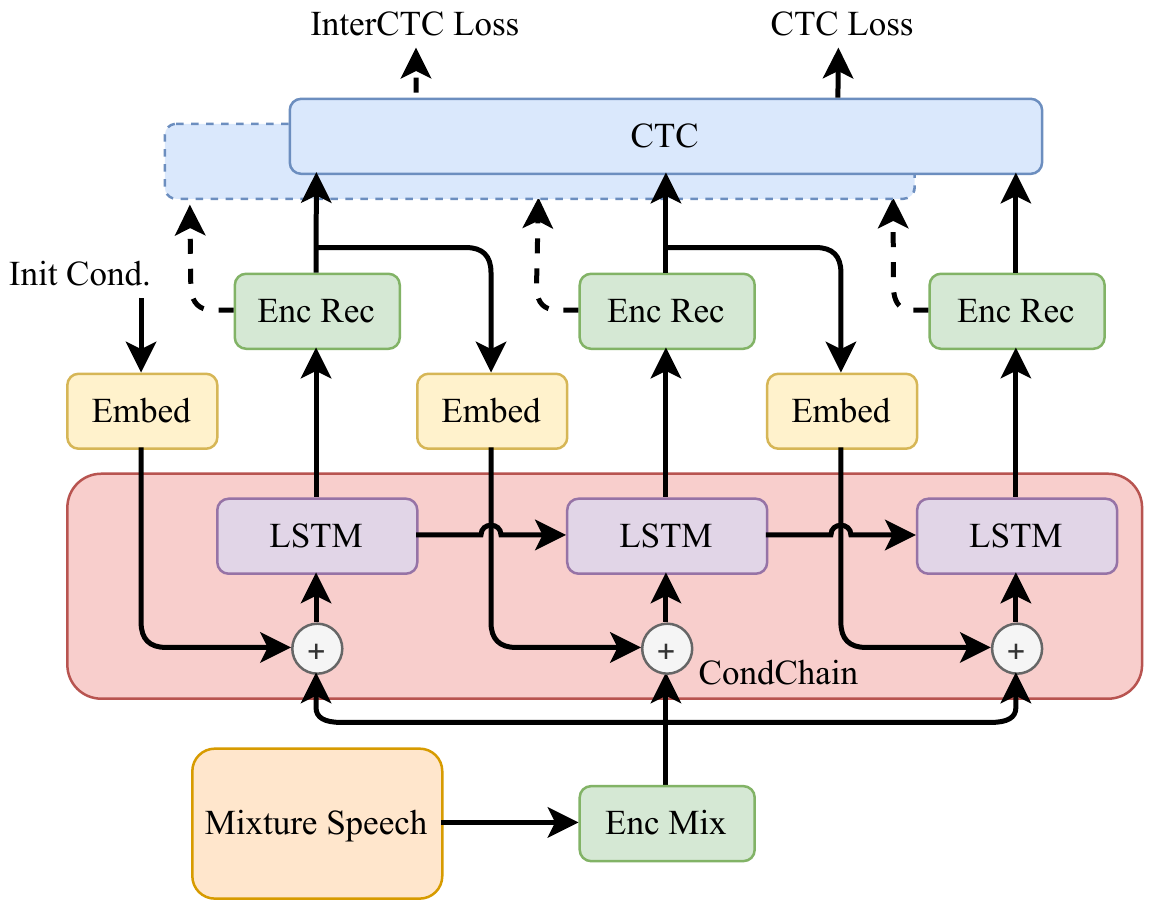}
    \caption{A overview of proposed conditional speaker chain based Conformer CTC model for NAR multi-speaker ASR. This figure shows a training procedure of a 3-speaker mixed waveform. The parameters of blocks with the same name are shared.}
    \label{fig:conditional-chain}
\end{figure}

\subsection{Conditional Chain Model}
Fig.~\ref{fig:conditional-chain} shows an overview of our model. Different from the AR models described in Section~\ref{sec:multispkr-asr}, we replace the SD encoders with a conditional speaker chain module (CondChain) and predict the output of each speaker one-by-one. With a hidden mixture representation $\mathbf{H}$ computed in Eq.~(\ref{eq:enc_mix}), the CondChain module extracts each speaker's speech representation by taking advantage of both the mixture representation $\mathbf{H}$ and the previously-estimated high-level embedding $\mathbf{G}^{j-1}$:
\begin{align}
    \mathbf{H}^{j} &= \text{CondChain}(\mathbf{H}, \text{Embed}(\mathbf{G}^{j-1})),\, j=1,\dots,J, \label{eq:cond_chain} 
\end{align}
where $\mathbf{G}^{j-1}$, obtaining from the $\text{Encoder}_{\text{Rec}}$ output for previous speaker, can be viewed as the speaker condition. The $\text{Embed}$ module is a multi-layer fully connect layer aiming to project the linguistic sequence $\mathbf{G}^{j-1}$ into the acoustic sub-space. In the first step, an all-zero vector will be initialized as the speaker condition. Besides, the long short-term memory (LSTM) layer also helps to provide all historic speaker conditions by the flowing states. With this design, we can successfully perform a NAR computation in each step and the total inference steps is a constant number equaling to the number of mixed speakers. Moreover, compared with other multi-speaker ASR methods, which have to fix the number of mixed speakers in the training data, our model can handle variable mixed data and further improve the performance. Algorithm~\ref{alg:training_algorithm} outlines the training procedure of our proposed model.


\begin{algorithm}[t]
    \caption{Training procedure of our model}
    \label{alg:training_algorithm}
    \begin{algorithmic}[1]
    \STATE {Initialize the model parameters $\pmb{\theta}$ and a all-zero condition $\mathbf{G}^{0}$ for the first step}
    \STATE {
    Given hyper parameters
    \begin{itemize}
      \item learning rate $\alpha$, InterCTC loss weight $\lambda$
    \end{itemize}
    }
    \STATE {Loading pre-trained model or not}
    \WHILE{Epoch $<$ TotalEpoch}
    \STATE Given the input features $\mathbf{X} = \{ \mathbf{x}_1, \dots, \mathbf{x}_T \}$ of a mixture speech and the corresponding transcriptions $\mathbf{Y} = \{ \mathbf{Y}^1, \dots, \mathbf{Y}^J \}$ of $J$ different speakers 
    \STATE Forward the $\text{Encoder}_{\text{mix}}$ with $\mathbf{X}$ and obtain the mixture hidden representations of $\mathbf{H}$ using Eq.~(\ref{eq:enc_mix})
    
    \FOR{ ($i=1$; $i<J$; $i\mbox{++}$) }
    \STATE Concatenate the $\textbf{H}$ with previously-estimated condition $\mathbf{G}^{i-1}$ and forward the LSTM layer as in Eq.~(\ref{eq:cond_chain})
    \vspace{-0.3cm}
    \STATE Forward the $\text{Encoder}_{\text{rec}}$ with the output of LSTM layer
    \vspace{-0.3cm}
    \STATE The output of $\text{Encoder}_{\text{rec}}$ is used to compute the $\text{Loss}_{\text{CTC}}$ as well as determine the best permutation of transcriptions as in Eq.~(\ref{eq:pit-ctc})
    \STATE The output of the intermediate layer in $\text{Encoder}_{\text{rec}}$ is used to compute the $\text{Loss}_{\text{InterCTC}}$ with above best transcription permutations
    \STATE $\mathbf{G}^{i}$ will also be regarded as the condition for the prediction of the next speaker
    \ENDFOR
    
    \STATE Update the model using Eq.~(\ref{eq:our_tot_loss})
    \STATE Epoch = Epoch + 1
    \ENDWHILE
    \RETURN $\pmb{\theta} $
    \end{algorithmic}
\end{algorithm}
\setlength{\textfloatsep}{0.4cm}

\section{Experiments}
\label{sec:experiments}

\subsection{Setup}
\label{ssec:setup}
The proposed models are evaluated on two commonly used simulated multi-speaker speech datasets.\\
\textbf{WSJ0-Mix.}~The dataset can be divided into two categories, namely the 2-speaker scenario and 3-speaker scenario. In the 2-speaker scenario, we use the common benchmark called WSJ0-2mix dataset introduced by~\cite{hershey2016deep} with a sampling rate of $16$~KHz. The training and validation sets are generated by randomly selecting two utterances from different speakers from the WSJ0 si\_tr\_s partition, containing around $30$~h and $10$~h speech mixture, respectively. To mix the utterances, various signal-to-noise ratios (SNRs) are uniformly chosen from [0, 10] dB. For the test set, the mixture is similarly generated using utterances from the WSJ0 validation set si\_dt\_05 and evaluation set si\_et\_05, resulting in $5$~h speech mixtures. For the 3-speaker experiments, similar methods are adopted except the number of speakers is three. \\
\textbf{LibriMix.}~Our methods are additionally tested on LibriMix, a recent open-source dataset for multi-speaker speech processing. The LibriMix data is created by mixing the source utterances randomly chosen from different speakers in LibriSpeech~\cite{panayotov2015librispeech} and the noise samples from WHAM!~\cite{wichern2019wham}. The SNRs of the mixtures are normally distributed with a mean of $0$ dB and a standard deviation of $4.1$ dB. LibriMix is composed of 2-speaker or 3-speaker mixtures, with or without noise conditions. For fast evaluation, we conducted our experiments on the $\text{train-}100$ subset from Libri2Mix, which contains around $100$~h of 2-speaker mixture speech.

All the proposed models are implemented with ESPnet~\cite{watanabe2018espnet}. 
We followed the ESPnet recipe to set the hyper-parameters of the model. For all Transformer- and Conformer-based models, EncoderMix is comprised of two CNN blocks and EncoderRec contains 8 Transformer or Conformer layers, depending on the model choices. For non-conditional chain models, the EncoderSD is a 4-layer Transformer or Conformer network, while the CondChain is a 1-layer LSTM network with 1024 hidden units. The common parameters of the Transformer and Conformer layers are: $d^{\text{head}}=4, d^{\text{att}}=256, d^{\text{ff}}=2048$ for the number of heads, dimension of attention module, and dimension of feed-forward layer, respectively.

\begin{table*}[tbp]
\centering
\caption{Word error rates (WERs) and real time factor (RTF) for multi-speaker speech recognition on WSJ0-Mix dataset. The RTF results are obtained by averaging the results of 5 decoding processes on CPUs.}
\label{table:exp-wsj0}
\begin{threeparttable}
\begin{tabular}{lcccc} 
\toprule
\multirow{2}{*}{\textbf{Models}} & \multirow{2}{*}{\textbf{Training Data}} & \multicolumn{2}{c}{\textbf{WER (\%)}} & \multirow{2}{*}{\textbf{RTF}}\\ 
& & \textbf{WSJ0-2mix} & \textbf{WSJ0-3mix} & \\ 
\hline\hline
\multicolumn{5}{l}{\textit{Hybrid model (w/ beam search)}} \\ 
\midrule
(1) PIT-DNN-HMM~\cite{qian2018single} & WSJ0-2mix & 28.2 & - &  - \\
(2) DPCL + DNN-HMM~\cite{menne2019analysis} & WSJ0-2mix & 16.5 & - & - \\
\hline\hline
\multicolumn{5}{l}{\textit{E2E Autoregressive Model (w/ greedy search)}} \\ 
\midrule
(3) PIT-RNN~\cite{chang2019end}\tnote{$\dagger$} & WSJ0-2mix & 51.4 & - & 1.4293 \\
(4) PIT-Transformer~\cite{shi2020sequence}\tnote{$\dagger$} & WSJ0-2mix & 37.0  & - & 1.4695 \\
(5) PIT-Conformer & WSJ0-2mix & 22.4 & - & 1.3970 \\ 
\hline\hline
\multicolumn{5}{l}{\textit{E2E Non-autoregressive Model (w/ greedy search)}} \\ 
\midrule
(6) PIT-Transformer-CTC & WSJ0-2mix & 50.3 & - & 0.1091 \\
(7) Conditional-Transformer-CTC~\cite{shi2020sequence}\tnote{$\dagger$} & WSJ0-2mix & 41.0 & - & 0.1293 \\
(8) Conditional-Transformer-CTC~\cite{shi2020sequence}\tnote{$\dagger$} & WSJ0-1\&2\&3mix & 29.4 & 53.3 & - \\
(9) Conditional-Conformer-CTC & WSJ0-2mix & 25.3 & - & 0.1824 \\
~ ~ + hidden feature conditions & WSJ0-2mix & 24.4 & - & 0.1758 \\
~ ~ + InterCTC loss & WSJ0-2mix & 22.3 & - & 0.1854 \\
(10) Conditional-Conformer-CTC & WSJ0-1\&2\&3mix & 23.4 & 39.1 & 0.1771 / 0.2096\\
~ ~ + hidden feature conditions & WSJ0-1\&2\&3mix & 22.2 & 38.6 & 0.1741 / 0.2241\\
~ ~ + InterCTC loss & WSJ0-1\&2\&3mix & \textbf{19.9} & \textbf{34.3} & 0.1732 / 0.2088\\
\bottomrule
\end{tabular}
\begin{tablenotes}
    \footnotesize
    \item $\dagger$: The results are obtained by the same implementation in~\cite{shi2020sequence} but w/o beam search and LM rescoring. When using both beam search and LM rescoring, the results are 14.9\% / 37.9\% of model (8) and 12.4\% / 26.6\% of model (10).
\end{tablenotes}
\end{threeparttable}
\vspace{-0.4cm}
\end{table*}

\begin{table}[t]
\centering
\caption{Word error rates (WERs) for multi-speaker speech recognition on LibriMix dataset.}
\label{table:exp-libri}
\begin{tabular}{lcc} 
\toprule
\textbf{\textbf{Models}}                   & \textbf{Dev} & \textbf{Test}  \\ 
\hline\hline
\multicolumn{3}{l}{\textit{E2E Autoregressive Model (w/ greedy search)}}        \\ 
\midrule
(1) PIT-Transformer & 34.8 & 36.0 \\
\hline\hline
\multicolumn{3}{l}{\textit{E2E Non-autoregressive Model (w/ greedy search)}}    \\ 
\midrule
(2) PIT-Transformer-CTC & 45.2 & 45.9 \\
(3) Conditional-Transformer-CTC & 32.7 & 33.3 \\
(4) Conditional-Conformer-CTC + both & \textbf{24.5} & \textbf{24.9} \\
\bottomrule
\end{tabular}
\end{table}

\subsection{Results on WSJ0-Mix} \label{sub:wsj0_mix}
In this part, we present the performance on the WSJ0-Mix corpus, which is shown in Table~\ref{table:exp-wsj0}. To evaluate the effectiveness, we compare our conditional speaker chain based Conformer CTC model with a variety of systems including the hybrid systems, PIT-based end-to-end AR and NAR models, and conditional speaker chain based Transformer models. Since all PIT-based models are unable to deal with variable numbers of speakers, only the results of 2-speaker scenario are presented. To make a fair comparison with NAR methods, the end-to-end AR models are decoded only with greedy search.

For the PIT-based AR models, PIT-Conformer (5) shows the best performance, achieving a word error rate (WER) of 22.4\% on the WSJ0-2mix test set. When comparing the NAR models, PIT-Transformer-CTC (6), which is only trained with CTC loss, suffers a dramatic performance degradation (50.3\%). There is no doubt that a pure CTC based encoder network can hardly model different speaker's speech simultaneously. When applying the conditional speaker chain based method, both model (7) and model (8) are better than PIT model. By combining the single and multi-speaker mixture speech, model (8) shows a significant improvement, whose WER is 29.5\% on the WSJ0-2mix test set.
For our conditional Conformer-CTC model (9), we explore two types of conditional features, including the ``hard" CTC alignments and ``soft" latent features after $\text{Encoder}_{\text{Rec}}$. Both approaches are better than above models with only a $\sim$0.07 seconds increase of latency and applying the ``soft" features achieves a WER of 24.4\%. By incorporating the intermediate loss, we can obtain a superior WER of 22.3\%, reaching a strong AR PIT-Conformer model (5). However, after combining latent feature conditions and the intermediate CTC loss, we don't get a further improvement.
Finally, we also train our model on the data of variable numbers of speakers and obtain the best WERs of 19.9\% and 34.3\%, which are even better than model (5) with only 1/7 latency.

\begin{table}[t]
\centering
\caption{Correlation between the hypothesis (Hyp.) generation order and the source signal (Src.) length order on WSJ0-2mix.}
\label{table:correlation}
\begin{tabular}{c|c|c}
    \toprule
    \diagbox{\textbf{Hyp.}}{\textbf{Src.}} & \begin{tabular}[c]{@{}c@{}} long \end{tabular} & \begin{tabular}[c]{@{}c@{}} short \end{tabular} \\ \hline\hline
    \begin{tabular}[c]{@{}c@{}}$1^{\text{st}}$ output \end{tabular} & 2749 & 251 \\ \hline
    \begin{tabular}[c]{@{}c@{}}$2^{\text{nd}}$ output \end{tabular} & 251 & 2749 \\ \bottomrule
\end{tabular}
\end{table}

We further investigate the correlation between the hypothesis generation order and the source signal length (from long to short), as shown in Table~\ref{table:correlation}. We find that only 251/3000 utterances do not follow the order on 2-speaker scenario and the average Spearman's Coefficient is $0.833$.



\subsection{Results on LibriMix} \label{sub:librimix}
The results on LibriMix are summarized in Table~\ref{table:exp-libri}. From the table, we can see a quite similar trend as the WSJ0-Mix results in the previous subsection. Our Conditional-Conformer-CTC with both latent features conditions and intermediate CTC loss obtains the best WERs of 24.5\% and 24.9\% on dev and test sets, respectively, which yields up to 25\% relative improvement compared with the Conditional-Transformer-CTC model.

\section{Conclusions} \label{sec:conclusion}
In this study, we revisit our proposed conditional speaker chain based multi-speaker ASR by enhancing the NAR ability. Our improved model mainly includes a conditional speaker chain (CondChain) module and Conformer CTC based encoders. To boost the performance of a pure Conformer CTC encoder, we also investigate two approaches, which are using the ``soft" latent features from the encoder output as speaker conditions and including an additional intermediate CTC loss. We evaluate the effectiveness of our model on two multi-speaker benchmarks, WSJ0-Mix and LibriMix. Our model shows consistent improvement over other models with only a slight increment of RTF and even better than a strong AR model in some cases. 



\clearpage
\bibliographystyle{IEEEtran}

\bibliography{mybib}


\end{document}